\documentclass[a4paper,conference]{IEEEtran}

\usepackage{amsfonts}
\usepackage{etex}
\usepackage{amsmath,amssymb}
\usepackage{mathtools,hyperref}
\usepackage{amsthm}
\usepackage{float}
\usepackage{slashbox}
\usepackage{subcaption}
\usepackage{nicefrac}
\usepackage[dvipsnames]{xcolor}
\usepackage{tikz}
\usepackage{ctable}
\usepackage{graphicx}  
\usepackage{tabularx}
\usepackage{accents}
\usepackage{enumitem}
\usepackage[makeroom]{cancel}
\usepackage{colortbl}
\usepackage{multirow,geometry,balance}
\theoremstyle{remark}
\newtheorem{theorem}{ {Theorem}}

\newtheorem{definition}{{Definition}}

\newtheorem{lemma}{ {Lemma}}
\newtheorem{remark}{ {Remark}}

\let\olddefinition\definition
\renewcommand{\definition}{\olddefinition\normalfont}
\let\oldtheorem\theorem
\renewcommand{\theorem}{\oldtheorem\normalfont}
\let\oldremark\remark
\renewcommand{\remark}{\oldremark\normalfont}
\usepackage{mathtools}
\usepackage{pgfplots}
\pgfplotsset{compat=newest}
\usetikzlibrary{plotmarks}
\usepackage{grffile}
\usepackage{amsmath}

\usepackage{balance}

\usetikzlibrary{patterns, arrows,backgrounds,fit,tikzmark,positioning,calc,decorations,snakes,shapes}

\newcommand{\ti}{\textit}

\renewcommand{\IEEEQED}{\IEEEQEDopen}
\DeclarePairedDelimiter\ceil{\lceil}{\rceil}

\usepackage{ellipsis}
\usetikzlibrary{calc}
\usetikzlibrary{decorations.pathreplacing,decorations.markings,shapes.geometric}
\tikzset{naming/.style={align=center,font=\small}}
\tikzset{antenna/.style={insert path={-- coordinate (ant#1) ++(0,0.25) -- +(135:0.25) + (0,0) -- +(45:0.25)}}}
\tikzset{station/.style={naming,draw,shape=dart,shape border rotate=90, minimum width=10mm, minimum height=10mm,outer sep=0pt,inner sep=3pt}}
\tikzset{mobile/.style={naming,draw,shape=rectangle,minimum width=12mm,minimum height=6mm, outer sep=0pt,inner sep=3pt}}
\tikzset{radiation/.style={{decorate,decoration={expanding waves,angle=90,segment length=4pt}}}}

\newenvironment{customlegend}[1][]{%
    \begingroup
    \csname pgfplots@init@cleared@structures\endcsname
    \pgfplotsset{#1}%
}{%
    \csname pgfplots@createlegend\endcsname
    \endgroup
}%

\def\addlegendimage{\csname pgfplots@addlegendimage\endcsname}

\newcommand{\SymMod}[0]{
\begin{tikzpicture}[
roundnode/.style={circle, draw=green!60, fill=green!5, very thick, minimum size=7mm},
squarednode/.style={rectangle, draw=red!60, fill=red!5, very thick, minimum size=5mm},
]
\node [cloud, draw,cloud puffs=10,cloud puff arc=120, aspect=2, inner ysep=1em, scale=0.7] (c) at (0, 2.6) {Cloud};
\node[squarednode] (eNB) at (2.0, 2.6) {DeNB};
\node[squarednode] (HeNB) at (3.75, 0.6) {RN};
\draw[->, thick] (c) -- (eNB);
\node[squarednode] (Utwo) at (5.5, 2.6) {UE};
\draw[->, dashed, thick] (2.6, 2.6) -- (5.1, 2.6) node[pos=0.5,sloped,above] {$h_{d}/n_{d}$};
\draw[->, dashed, thick] (2.6, 2.6) -- (3.3, 0.6) node[pos=0.5,sloped,below] {$h_{s}/n_{s}$};
\draw[->, dashed, thick] (4.15, 0.6) -- (5.1, 2.5) node[pos=0.5,sloped,below] {$h_{r}/n_{r}$};
\node (ch) at (3.75, 0) {\includegraphics[scale=.3]{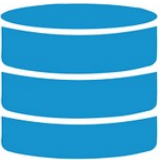}};
\draw[<->] (4.1, -0.23) -- (4.1, 0.24) node[pos=0.5,sloped, below] {$\scriptstyle\mu NL$};
\end{tikzpicture}
}

\newlength{\stackWidth}
\newlength{\smallestHeight}
\newlength{\stackWidthTwo}
\newlength{\smallestHeightTwo}
\newlength{\smalleps}
\tikzset{block1/.style={shape=rectangle, draw, node distance=-1pt, minimum width = \stackWidth, line width=1pt, anchor=north west, inner sep=0pt}}
\tikzset{block2/.style={shape=rectangle, draw, node distance=-1pt, minimum width = \stackWidthTwo, line width=1pt, anchor=mid west, inner sep=0pt}}

\newcommand{\Achievability}[0]{
\coordinate (DeNBupleft) at (0,2);
\coordinate (RNupleft) at (2.5,0);
\coordinate (UEupleft) at (6.2,2);

\coordinate (filesleft) at (2.2,2.5);
\node (W1) at (filesleft) {$W_1$}; 
\begin{scope}[every node/.style={block2}]
\node[minimum height=\smallestHeightTwo,minimum width=3.125\stackWidthTwo,fill=green!20] at ([xshift=0.35cm, yshift=0.1cm]filesleft) (f1) {\tiny $W_1^{\bar{c}}$};
\node[minimum height=\smallestHeightTwo,minimum width=1.875\stackWidthTwo, fill=green!50, right=of f1] (f2) {\tiny $W_1^{c}$};
\end{scope}
\node (W2) at ([yshift=-0.5cm]W1) {$W_2$}; 
\begin{scope}[every node/.style={block2}]
\node[minimum height=\smallestHeightTwo,minimum width=3.125\stackWidthTwo,fill=yellow!20] at ([xshift=0.35cm, yshift=0.1cm]W2) (g1) {\tiny $W_2^{\bar{c}}$};
\node[minimum height=\smallestHeightTwo,minimum width=1.875\stackWidthTwo, fill=yellow!50, right=of g1] (g2) {\tiny $W_2^{c}$};
\end{scope}

\node (lefttopcorner) at ([xshift=0.35cm+3.125\stackWidthTwo-\smalleps, yshift=0.1cm+0.5\smallestHeightTwo+\smalleps]filesleft) {};
\node (righttopcorner) at ([xshift=0.35cm+5\stackWidthTwo+\smalleps, yshift=0.1cm+0.5\smallestHeightTwo+\smalleps]filesleft) {};
\node (leftbottomcorner) at ([yshift=-0.4cm-1.7\smallestHeightTwo-\smalleps]lefttopcorner) {};
\node (rightbottomcorner) at ([yshift=-0.4cm-1.7\smallestHeightTwo-\smalleps]righttopcorner) {};
\draw[red,very thick,dashed] (lefttopcorner)--(righttopcorner)--(rightbottomcorner)--(leftbottomcorner)--(lefttopcorner);
\node (text) at ([xshift=0.9375\stackWidthTwo+\smalleps, yshift=5\smalleps]lefttopcorner) {\footnotesize Cached Bits};

\node (DeNB) at ([xshift=0.5\stackWidth]DeNBupleft) {DeNB}; 
\begin{scope}[every node/.style={block1}]
\node[minimum height=2\smallestHeight,fill=yellow!20] at ([yshift=-0.2cm]DeNBupleft) (b1) {$W_2^{\bar{c}}$};
\node[minimum height=2\smallestHeight,fill=green!20, below=of b1] (b2) {$W_1^{\bar{c}}$};
\node[minimum height=\smallestHeight,below=of b2] (b3) {\footnotesize $\mathbf{0}_{n_r-n_s}$};\end{scope}

\node (BlockTopLeft) at ([yshift=-0.2cm]DeNBupleft) {}; 
\node (BlockTopRight) at ([xshift=\stackWidth]BlockTopLeft) {}; 
\node (b1right) at ([yshift=-2\smallestHeight]BlockTopRight){};
\node (b2right) at ([yshift=-4\smallestHeight]BlockTopRight){};
\node (b3right) at ([yshift=-5\smallestHeight]BlockTopRight){};

\draw [decorate,decoration={brace,amplitude=3.5pt,raise=4pt},yshift=0pt]
(BlockTopRight) -- (b1right) node [black,midway,xshift=0.65cm] {\footnotesize $\nicefrac{n_s}{2}$};
\draw [decorate,decoration={brace,amplitude=3.5pt,raise=4pt},yshift=0pt]
(b1right) -- (b2right) node [black,midway,xshift=0.65cm] {\footnotesize $\nicefrac{n_s}{2}$};

\node (RN) at ([xshift=\stackWidth+0.1cm]RNupleft) {RN}; 
\begin{scope}[every node/.style={block1}]
\node[minimum height=\smallestHeight] at ([yshift=-0.2cm]RNupleft) (r1) {\footnotesize $\mathbf{0}_{n_r-n_s}$};
\node[minimum height=2\smallestHeight,fill=yellow!20, below=of r1] (r2) {$W_2^{\bar{c}}$};
\node[minimum height=2\smallestHeight,fill=green!20, below=of r2] (r3) {$W_1^{\bar{c}}$};
\node[anchor=north west, minimum height=1.5\smallestHeight, fill=yellow!50, below right=0cm and 0.1cm of r1.north east] (r4) {$W_2^c$};
\node[minimum height=3.5\smallestHeight, below=of r4] (r5) {\footnotesize $\mathbf{0}_{n_d}$};
\end{scope}

\node (RNBlockTopLeft) at ([yshift=-0.2cm]RNupleft) {};
\node (RNBlockTopRight) at ([xshift=2\stackWidth+0.1cm]RNBlockTopLeft) {};
\node (r1right) at ([yshift=-1.5\smallestHeight]RNBlockTopRight){};
\draw [decorate,decoration={brace,amplitude=2.5pt,raise=4pt},yshift=0pt]
(RNBlockTopRight) -- (r1right) node [black,midway,xshift=0.80cm] {\footnotesize $n_r-n_d$};
\node (RNBlockBottomLeftMiddle) at ([yshift=-5\smallestHeight-0.2cm, xshift=0.5\stackWidth]RNBlockTopLeft) {\footnotesize received};
\node (RNBlockBottomRightMiddle) at ([yshift=-5\smallestHeight-0.2cm, xshift=-0.5\stackWidth+0.05cm]RNBlockTopRight) {\footnotesize transmitted};

\node (UE) at ([xshift=0.5\stackWidth]UEupleft) {UE}; 
\begin{scope}[every node/.style={block1}]
\node[minimum height=1.5\smallestHeight,fill=yellow!50] at ([yshift=-0.2cm]UEupleft) (u1) {$W_2^{c}$};
\node[minimum height=2\smallestHeight,fill=yellow!20, below=of u1] (u2) {$W_2^{\bar{c}}$};
\node[minimum height=1.5\smallestHeight,fill=green!20, below=of u2] (u3) {};\end{scope}

\node (UEBlockTopLeft) at ([yshift=-0.2cm]UEupleft) {}; 
\node (UEBlockTopRight) at ([xshift=\stackWidth]UEBlockTopLeft) {}; 
\node (u1right) at ([yshift=-3.5\smallestHeight]UEBlockTopRight){};
\node (u2right) at ([yshift=-5\smallestHeight]UEBlockTopRight){};

\draw [decorate,decoration={brace,amplitude=2.5pt,raise=4pt},yshift=0pt]
(u1right) -- (u2right) node [black,midway,xshift=0.95cm] {\footnotesize $n_d-\nicefrac{n_s}{2}$};

\node (BlockMiddleRight) at ([yshift=-2.1\smallestHeight]BlockTopRight) {}; 
\node (BlockMiddleBottom) at ([xshift=0.5\stackWidth,yshift=-5\smallestHeight]BlockTopLeft) {}; 

\node (BlockRNMiddleLeft) at ([yshift=-3.55\smallestHeight]RNBlockTopLeft) {}; 
\node (BlockRNMiddleRight) at ([yshift=-3.55\smallestHeight]RNBlockTopRight) {}; 

\node (BlockUEMiddleLeft) at ([yshift=-2.1\smallestHeight]UEBlockTopLeft) {}; 
\node (BlockUEMiddleBottom) at ([xshift=0.5\stackWidth,yshift=-5\smallestHeight]UEBlockTopLeft) {}; 

\draw (BlockMiddleBottom) edge[bend right,->] node [below] {$n_s$}(BlockRNMiddleLeft);
\draw (BlockRNMiddleRight) edge[bend right,->] node [below, pos=0.35] {$n_r$}(BlockUEMiddleBottom);
\draw[->] (BlockMiddleRight)--(BlockUEMiddleLeft) node [above,pos=0.5] {$n_d\geq\nicefrac{n_s}{2}$};
}

\newcommand{\AchievabilityTwo}[0]{
\coordinate (DeNBupleft) at (0,2);
\coordinate (RNupleft) at (2.5,0);
\coordinate (UEupleft) at (6.2,2);

\coordinate (filesleft) at (2.2,2.5);
\node (W1) at (filesleft) {$W_1$}; 
\begin{scope}[every node/.style={block2}]
\node[minimum height=\smallestHeightTwo,minimum width=1.45\stackWidthTwo,fill=green!20] at ([xshift=0.35cm, yshift=0.1cm]filesleft) (f1) {\tiny $W_1^{\bar{c}}$};
\node[minimum height=\smallestHeightTwo,minimum width=3.55\stackWidthTwo, fill=green!50, right=of f1] (f2) {\tiny $W_1^{c}$};
\end{scope}
\node (W2) at ([yshift=-0.5cm]W1) {$W_2$}; 
\begin{scope}[every node/.style={block2}]
\node[minimum height=\smallestHeightTwo,minimum width=1.45\stackWidthTwo,fill=yellow!20] at ([xshift=0.35cm, yshift=0.1cm]W2) (g1) {\tiny $W_2^{\bar{c}}$};
\node[minimum height=\smallestHeightTwo,minimum width=3.55\stackWidthTwo, fill=yellow!50, right=of g1] (g2) {\tiny $W_2^{c}$};
\end{scope}

\node (lefttopcorner) at ([xshift=0.35cm+1.45\stackWidthTwo-\smalleps, yshift=0.1cm+0.5\smallestHeightTwo+\smalleps]filesleft) {};
\node (righttopcorner) at ([xshift=0.35cm+5\stackWidthTwo+\smalleps, yshift=0.1cm+0.5\smallestHeightTwo+\smalleps]filesleft) {};
\node (leftbottomcorner) at ([yshift=-0.4cm-1.7\smallestHeightTwo-\smalleps]lefttopcorner) {};
\node (rightbottomcorner) at ([yshift=-0.4cm-1.7\smallestHeightTwo-\smalleps]righttopcorner) {};
\draw[red,very thick,dashed] (lefttopcorner)--(righttopcorner)--(rightbottomcorner)--(leftbottomcorner)--(lefttopcorner);
\node (text) at ([xshift=1.775\stackWidthTwo+\smalleps, yshift=5\smalleps]lefttopcorner) {\footnotesize Cached Bits};

\node (DeNB) at ([xshift=0.5\stackWidth]DeNBupleft) {DeNB}; 
\begin{scope}[every node/.style={block1}]
\node[minimum height=1.1\smallestHeight] at ([yshift=-0.2cm]DeNBupleft) (b1) {\footnotesize $\mathbf{0}_{2n_d-n_s}$};
\node[minimum height=1.45\smallestHeight,fill=yellow!20, below=of b1] (b2) {$W_2^{\bar{c}}$};
\node[minimum height=1.45\smallestHeight,fill=green!20, below=of b2] (b3) {$W_1^{\bar{c}}$};
\node[minimum height=\smallestHeight,below=of b3] (b4) {\footnotesize $\mathbf{0}_{n_r-n_s}$};\end{scope}

\node (BlockTopLeft) at ([yshift=-0.2cm]DeNBupleft) {}; 
\node (BlockTopRight) at ([xshift=\stackWidth]BlockTopLeft) {}; 
\node (b1right) at ([yshift=-1.1\smallestHeight]BlockTopRight){};
\node (b2right) at ([yshift=-2.55\smallestHeight]BlockTopRight){};
\node (b3right) at ([yshift=-4\smallestHeight]BlockTopRight){};

\draw [decorate,decoration={brace,amplitude=2pt,raise=4pt},yshift=0pt]
(b1right) -- (b2right) node [black,midway,xshift=0.85cm] {\footnotesize $n_s-n_d$};
\draw [decorate,decoration={brace,amplitude=2pt,raise=4pt},yshift=0pt]
(b2right) -- (b3right) node [black,midway,xshift=0.85cm] {\footnotesize $n_s-n_d$};

\node (RN) at ([xshift=\stackWidth+0.1cm]RNupleft) {RN}; 
\begin{scope}[every node/.style={block1}]
\node[minimum height=\smallestHeight] at ([yshift=-0.2cm]RNupleft) (r1) {\footnotesize $\mathbf{0}_{n_r-n_s}$};
\node[minimum height=1.1\smallestHeight,below=of r1] (r2) {\footnotesize $\mathbf{0}_{2n_d-n_s}$};
\node[minimum height=1.45\smallestHeight,fill=yellow!20, below=of r2] (r3) {$W_2^{\bar{c}}$};
\node[minimum height=1.45\smallestHeight,fill=green!20, below=of r3] (r4) {$W_1^{\bar{c}}$};
\node[anchor=north west, minimum height=3.55\smallestHeight, fill=yellow!50, below right=0cm and 0.1cm of r1.north east] (r4) {$W_2^c$};
\node[minimum height=1.45\smallestHeight, below=of r4] (r5) {\footnotesize $\mathbf{0}_{n_s-n_d}$};
\end{scope}

\node (RNBlockTopLeft) at ([yshift=-0.2cm]RNupleft) {};
\node (RNBlockTopRight) at ([xshift=2\stackWidth+0.1cm]RNBlockTopLeft) {};
\node (r1right) at ([yshift=-3.55\smallestHeight]RNBlockTopRight){};
\draw [decorate,decoration={brace,amplitude=3.5pt,raise=4pt},yshift=0pt]
(RNBlockTopRight) -- (r1right) node [black,midway,xshift=0.85cm, yshift=0.35cm, align=left] {\footnotesize $n_d+n_r$ \\ \footnotesize $-n_s$};
\node (RNBlockBottomLeftMiddle) at ([yshift=-5\smallestHeight-0.2cm, xshift=0.5\stackWidth]RNBlockTopLeft) {\footnotesize received};
\node (RNBlockBottomRightMiddle) at ([yshift=-5\smallestHeight-0.2cm, xshift=-0.5\stackWidth+0.05cm]RNBlockTopRight) {\footnotesize transmitted};

\node (RN) at ([xshift=0.5\stackWidth]UEupleft) {UE}; 
\begin{scope}[every node/.style={block1}]
\node[minimum height=3.55\smallestHeight,fill=yellow!50] at ([yshift=-0.2cm]UEupleft) (u1) {$W_2^{c}$};
\node[minimum height=1.45\smallestHeight,fill=yellow!20, below=of u1] (u2) {$W_2^{\bar{c}}$};
\end{scope}

\node (BlockMiddleRight) at ([yshift=-2.5\smallestHeight]BlockTopRight) {}; 
\node (BlockMiddleBottom) at ([xshift=0.5\stackWidth,yshift=-5\smallestHeight]BlockTopLeft) {}; 

\node (BlockRNMiddleLeft) at ([yshift=-3.55\smallestHeight]RNBlockTopLeft) {}; 
\node (BlockRNMiddleRight) at ([yshift=-3.55\smallestHeight]RNBlockTopRight) {}; 

\node (BlockUEMiddleLeft) at ([yshift=-2.5\smallestHeight]UEBlockTopLeft) {}; 
\node (BlockUEMiddleBottom) at ([xshift=0.5\stackWidth,yshift=-5\smallestHeight]UEBlockTopLeft) {}; 

\draw (BlockMiddleBottom) edge[bend right,->] node [below] {$n_s$}(BlockRNMiddleLeft);
\draw (BlockRNMiddleRight) edge[bend right,->] node [below, pos=0.35] {$n_r$}(BlockUEMiddleBottom);
\draw[->] (BlockMiddleRight)--(BlockUEMiddleLeft) node [above,pos=0.5] {$n_d$};
}

\newcommand*\circled[1]{\tikz[baseline=(char.base)]{
            \node[shape=circle,draw,inner sep=2pt] (char) {#1};}}

\newcommand{\Resultone}[0]{
\begin{axis}[
        xmin=-0.15,
        xlabel=$\mu$,
        ylabel={DTB -- ${\Delta^{*}_{\text{det}}(\mu,\mathbf{n}})$}, 					xtick={0, 0.6, 0.8333333333, 1},  
        ytick={0.5, 0.16666666666},
        extra y ticks={0.2},
        xticklabels={$0$,${\mu^{\prime}(\mathbf{n})}$,${\mu^{\prime\prime}(\mathbf{n})}$,$1$},  
        yticklabels={$\frac{2}{n_s}$, $\frac{1}{n_r}$}, 
        extra y tick labels={$\frac{1}{\nicefrac{n_{s}}{2}+n_r-n_{d}}\quad\:$},
        extra y tick style={%
            ,grid=major, tick label style={rotate=270},
            ,ticklabel pos=right
            },
        grid,
        legend cell align=left]
    
   \addplot[mark=*,blue, ultra thick] plot coordinates {
        (0,1/2) (3/5,1/5) (5/6,1/6) (1, 1/6)
    };
	\node[circle,fill,inner sep=2pt] (p1) at (axis cs:0,0.5) {};
	\node[circle,fill,inner sep=2pt] (p2) at (axis cs:0.6,0.2) {}; 
	\node[circle,fill,inner sep=2pt] (p3) at (axis cs:0.833333333,0.16666666) {}; 
	\node[circle,fill,inner sep=2pt] (p4) at (axis cs:1,0.16666666) {}; 
	\node (c1) at (axis cs: -0.075, 0.47) {\circled{$A$}}; 
	\node (c2) at (axis cs: 0.64, 0.23) {\circled{$B$}}; 
	\node (c3) at (axis cs: 0.8333333, 0.21) {\circled{$C$}};
	
	\node (rect) at (axis cs: 0.75, 0.38) [draw,thick,minimum width=1cm,minimum height=1.5cm, fill=white, align=left] {$\mu^{\prime}(\mathbf{n})=\frac{n_r-n_d}{\nicefrac{n_s}{2}+n_r-n_d}$ \\[1em] $\mu^{\prime\prime}(\mathbf{n})=\frac{n_d+n_r-n_s}{n_r}$};
\end{axis}
}

\allowdisplaybreaks 

\begin{document}

\title{Fundamental Limits on Latency in Transceiver Cache-Aided HetNets}

\author{\IEEEauthorblockN{Jaber Kakar, Soheil Gherekhloo and Aydin Sezgin}
\IEEEauthorblockA{Institute of Digital Communication Systems\\
Ruhr-Universit{\"a}t Bochum, 44780 Bochum, Germany\\
Email: \{jaber.kakar, soheyl.gherekhloo, aydin.sezgin\}@rub.de
}}

\maketitle

\begin{abstract}
Stringent mobile usage characteristics force wireless networks to undergo a paradigm shift from conventional connection-centric to content-centric deployment. With respect to 5G, caching and heterogenous networks (HetNet) are key technologies that will facilitate the evolution of highly content-centric networks by facilitating unified quality of service in terms of low-latency communication. In this paper, we study the impact of transceiver caching on the latency for a HetNet consisting of a single user, a receiver and one cache-assisted transceiver. We define an information-theoretic metric, the delivery time per bit (DTB), that captures the delivery latency. We establish coinciding lower and upper bounds on the DTB as a function of cache size and wireless channel parameters; thus, enabling a complete characterization of the DTB optimality of the network under study. As a result, we identify cache beneficial and non-beneficial channel regimes.  
\end{abstract} 

\IEEEpeerreviewmaketitle

\section{Introduction}
\label{sec:intro}

In recent years, mobile usage characteristics in wireless networks have changed profoundly from conventional connection-centric (e.g., phone calls) to content-centric (e.g, HD video) behaviors. As a result, content caching and heterogenous network (HetNet) technology are two major solutions for next generation (5G) mobile networks. Advantages of these solutions are two-fold: Firstly, caching the most popular contents in network edges, e.g., base stations and relays, alleviates backhaul traffic and reduces latency. Secondly, seamless uniform quality of service through improved performance, particularly, at cell edges is guaranteed. It is therefore to be expected that future networks will be heterogenous in nature, vastly deploying relay nodes (RN) (e.g., in LTE-A \cite{network_m2_2011}) endowed with (adaptive) cache capabilties. A simplistic HetNet modeling this aspect is shown in Fig. \ref{fig:HetNet}. In this model, the RN acts as a cache-aided transceiver. Thus, aspects of both transmitter and receiver caching in a single RN is captured through this network model enabling low-latency transmission and reception of requested files by RN and the user equipment (UE). In this work, we focus on characterizing the fundamental trade-off on latency of this particular network.

Various works showed that both receiver (Rx) and transmitter (Tx) caching offer great potential for reducing latency. Rx caching, on the one hand, was first studied in \cite{Maddah-Ali2} for a shared link with one server and multiple cache-enabled receivers. The authors show that caching can exploit multicast opportunities and as such significantly reduces the delivery latency over the shared link. On the other hand, the impact of Tx caching on the latency has mainly been investigated by analyzing the latency-centric metric; the inverse degrees-of-freedom (DoF), for Gaussian networks. To this end, the authors of \cite{Maddah_Ali} developed a novel achievability scheme characterizing the metric as a function of the cache storage capability for a 3-user Gaussian interference network. The cache placement was designed to facilitate transmitter cooperation such that interference coordination techniques can be applied. A converse on this metric was developed in \cite{avik} for a network with arbitrary number of edge nodes and users showing the optimality of schemes presented in \cite{Maddah_Ali} for certain regimes of cache sizes. Extensions of this work include the characterization of the latency-memory tradeoff to cloud and cache-assisted networks \cite{Tandon}.
Two new lines of research are to determine the fundamental limits of joint Tx-Rx caching at distinct nodes \cite{Naderializadeh} and transceiver caching. 
This paper focuses on the latter. 

In this paper, we are studying the fundamental limits on the latency for a \emph{transceiver} cache-aided HetNet consisting of a donor eNB (DeNB), a transceiver and a user. We measure the performance through the latency-centric metric \emph{delivery time per bit} (DTB). 
To this end, we establish coinciding lower (converse) and upper bounds (achievability) on the DTB for the linear deterministic model (LDM) \cite{Avestimehr} of our proposed network for various channel regimes. Hereby, the LDM serves as an approximation of the additive Gaussian noise model. 

\textbf{Notation:} For any two integers $a$ and $b$ with $a\leq b$, we define $[a:b]\triangleq\{a,a+1,\ldots,b\}$. When $a=1$, we simply write $[b]$ for $\{1,\ldots,b\}$. The superscript $(\cdot)^{\dagger}$ represents the transpose of a matrix. Furthermore, we define the function $(x)^{+}\triangleq\max\{0,x\}$. 
   
\section{System Model}
\label{sec:Sym_Model}
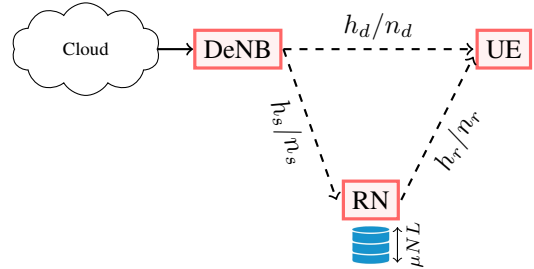
\begin{figure}[t]
        \begin{tikzpicture}[scale=0.8]
		\SymMod
		\end{tikzpicture}
	\caption{\small System model of cloud-aided HetNet for (a) Gaussian and (b) Linear Deterministic model}	
	\label{fig:HetNet}
\end{figure}
We study the downlink of a cache-aided HetNet as shown in Fig. \ref{fig:HetNet}. The HetNet consists of a causal full-duplex RN and a macro donor eNB (DeNB) which serve a single user -- a small-cell user (UE) -- over a wireless channel. Simultaneosuly, the RN also acts as a user requesting information from the DeNB. At every transmission interval, we assume that RN and UE request files from the set $\mathcal{W}$ of $N$ popular files, whose elements are all of $L$ bits in size. The transmission interval terminates when the requested files have been delivered. The system model, notation and main assumptions for a \emph{single} transmission interval are summarized as follows:
\begin{itemize}[leftmargin=0cm,itemindent=.2cm,labelwidth=\itemindent,labelsep=0cm,align=left]
\item Let $\mathcal{W}=\{W_1,\ldots,W_{N}\}$ denote the library of popular files, where each file $W_i$ is of size $L$ bits. Each file $W_i$ is chosen uniformly at random from $[2^{L}]$. RN and UE request files $W_{d_r}$ and $W_{d_u}$ from the library $\mathcal{W}$, respectively. The demand vector $\mathbf{d}=(d_r,d_u)^{\dagger}\in[N]^{2}$ denotes the request pattern of RN and UE.  
\item The RN is endowed with a cache capable of storing $\mu NL$ bits, where $\mu\in[0,1]$ corresponds to the fractional cache size. It denotes how much
content can be stored at the RN relatively to the entire library $\mathcal{W}$.
\item The cloud server has access to all $N$ files. The DeNB is connected to the cloud via a fronthaul link of infinite capacity $C_{F}=\infty$ bits per channel use. 
\item Global channel state information (CSI) for a single transmission interval is summarized by the channel vector $\mathbf{h}=(h_d,h_r,h_s)^{\dagger}\in\mathbb{C}^{3}$, where on the one hand $h_d$ and $h_s$ represent the complex channel coefficients from DeNB to the UE and RN, respectively. On the other hand, $h_r$ is the channel from RN to UE. We assume that all channel coefficients are assumed to be drawn i.i.d. from a continuous distribution and are kept fixed over a transmission interval.  
\end{itemize} 
Communication over the wireless channel occurs in two consecutive phases, \emph{placement phase} followed by the \emph{delivery phase}.  
In the following, we will describe the modeling of placement and delivery phase and the key performance metric termed as delivery time per bit (DTB) formally. 
\begin{enumerate}[leftmargin=0cm,itemindent=.4cm,labelwidth=\itemindent,labelsep=0cm,align=left]
\item \vspace{.5em} \emph{Placement phase}: During this phase, the RN is given full access to the database of $N$ files. The cached content at the RN is generated through its caching function. 
\vspace{.5em}
\begin{definition}(Caching function)\label{def_cache_fct} The RN maps each file $W_i\in\mathcal{W}$ to its local \emph{file cache content} 
\begin{equation}
S_{i}=\phi_{i}(W_{i}),\qquad\forall i=1,\ldots,N\nonumber.
\end{equation} 
All $S_{i}$ are concatenated to form the total cache content 
\begin{equation}
S=(S_{1},S_{2},\ldots,S_{N})\nonumber
\end{equation} 
at the RN. Hereby, due to the assumption of symmetry in caching, the entropy $H(S_{i})$ of each component $S_i$, $i=1,\ldots,N$, is upper bounded by $\nicefrac{\mu NL}{N}=\mu L$. The definition of the caching function presumes that every file $W_i$ is subjected to individual caching functions. Thus, permissible caching policies allow for intra-file coding but avoid inter-file coding. Moreover, the caching policy is typically kept fixed over multiple transmission intervals. Thus, it is indifferent to the user's request pattern and of channel realizations.     
\end{definition}
\item \vspace{.5em} \emph{Delivery phase}: In this phase, a transmission policy at DeNB and RN is applied to satisfy the given user's requests $\mathbf{d}$ under the current channel realizations $\mathbf{h}$. 
\vspace{.5em}
\begin{definition}(Encoding functions)\label{def_enc_fct} The DeNB encoding function 
\begin{equation}
\psi_{S}:[2^{NL}]\times [N]^{2}\times\mathbb{C}^{3}\rightarrow \mathbb{C}^{T}\nonumber
\end{equation} determines the transmission signal $\mathbf{x}_{s}^{T}$ subjected to an average power constraint of $P$. Hereby, the codeword $\mathbf{x}_{s}^{T}$ is a function of $\mathcal{W},\mathbf{d}$ and $\mathbf{h}$ conveyed to both RN and UE over $T$ channel uses. The encoding function of the causal full-duplex RN at the $t$--th time instant is defined by
\begin{equation}
\psi_{r}^{[t]}:[2^{\mu NL}]\times \mathbb{C}^{t-1}\times [N]^{2}\times\mathbb{C}^{3}\rightarrow \mathbb{C},\qquad t\in[1:T]\nonumber. 
\end{equation} For any time instant $t$, $\psi_{r}^{[t]}$ accounts for the simultaneous reception and transmission through incoming and outgoing wireless links at the RN. To be specific, at the $t$--th channel use the encoding function $\psi_{r}^{[t]}$ maps the cached content $S$, the received signal $\mathbf{y}_r^{t-1}$ (see Eq. \eqref{eq:Gaus_mod_RN}), the demand vector $\mathbf{d}$ and global CSI given by $\mathbf{h}$ to the codeword $\mathbf{x}_{r}[t]$ while satisfying the average power constraint given by the parameter $P$.
\end{definition}
\end{enumerate}
\vspace{.5em}
\begin{definition}(Decoding functions)\label{def_dec_fct} The decoding operation at the UE follows the mapping
\begin{equation}
\eta_{u}:\mathbb{C}^{T}\times [N]^{2}\times\mathbb{C}^{3}\rightarrow [2^{L}]\nonumber. 
\end{equation} The decoding function $\eta_u$ takes as its arguments $\mathbf{h}$, the available demand pattern $\mathbf{d}$ and the channel outputs $\mathbf{y}_{u}^{T}$ given by
\begin{equation}\label{eq:Gaus_mod}
\mathbf{y}_u^{T}=
h_{d}\mathbf{x}_{s}^{T}+h_{r}\mathbf{x}_{r}^{T}+\mathbf{z}_{u}^{T}
\end{equation} to provide an estimate $\hat{W}_{d_u}=\eta_{u}\big(\mathbf{y}_u^{T},\mathbf{d},\mathbf{h}\big)$ of the requested file $W_{d_u}$. The term $\mathbf{z}_{u}^{T}$ denotes complex i.i.d. Gaussian noise of zero mean and unit power. In contrast to decoding at the UE, the RN explicitly leverages its cached content according to
\begin{equation}
\eta_{r}:\mathbb{C}^{T}\times [2^{\mu NL}]\times [N]^{2}\times\mathbb{C}^{3}\rightarrow [2^{L}]\nonumber 
\end{equation} to generate an estimate $\hat{W}_{d_r}=\eta_{r}\big(\mathbf{y}_r^{T},S,\mathbf{d},\mathbf{h}\big)$ on the requested file $W_{d_r}$. Hereby, $\mathbf{y}_r^{T}$ is the received signal
\begin{equation}\label{eq:Gaus_mod_RN}
\mathbf{y}_r^{T}=
h_{s}\mathbf{x}_{s}^{T}+\mathbf{z}_{r}^{T}
\end{equation} corrupted through additive zero mean, unit-power i.i.d. Gaussian noise $\mathbf{z}_{r}^{T}$ at the RN in $T$ channel uses. 
\end{definition} A proper choice of a caching, encoding and decoding function that satisfies the reliability condition; that is, the worst-case error probability \begin{equation}\label{eq:error_prob}
P_e=\max_{\mathbf{d}\in [N]^{2}}\max_{j\in\{r,u\}}\mathbb{P}(\hat{W}_{d_j}\neq W_{d_j})
\end{equation} approaches $0$ as $L\rightarrow\infty$, is called a \emph{feasible policy}. 
\vspace{.5em}
\begin{definition}(Delivery time per bit) \cite{avik} The DTB for given request pattern $\mathbf{d}$ and channel realization $\mathbf{h}$ is defined as 
\begin{equation}\label{eq:DTB}
\Delta(\mu,\mathbf{h}, P)=\max_{\mathbf{d}\in [N]^{2}}\limsup_{L\rightarrow\infty}\frac{T(\mathbf{d},\mathbf{h})}{L}.
\end{equation} 
The minimum DTB $\Delta^{*}(\mu,\mathbf{h}, P)$ is the infimum of the DTB of all achievable schemes.
\end{definition}  
\begin{remark}
The DTB measures the per-bit latency, i.e., the latency incurred when transmitting the requested files through the wireless channel, within a single transmission interval for the \emph{worst-case} request pattern of RN and UE.
\end{remark}   
To gain initial insight into the DTB for the Gaussian system model, we suggest to approximate \eqref{eq:Gaus_mod} and \eqref{eq:Gaus_mod_RN} by the linear-deterministic model (LDM) \cite{Avestimehr}. In the LDM, input symbols at the DeNB and RN are given by the binary input vectors $\mathbf{x}_s$ and $\mathbf{x}_r\in\mathbb{F}_2^{q}$ where $q=\max\{n_{d},n_{r},n_{s}\}$. Hereby, the integers $n_{k}\in\mathbb{N}_{> 0}$, $k\in\{d,r,s\}$, given by
\begin{equation}\label{eq:nd}
n_{k}=\ceil{\log\big(P|h_k|^{2}\big)}
\end{equation} approximate the number of bits per channel use which can be communicated over each link reliably. The channel output symbols $\mathbf{y}_r^{T}$ and $\mathbf{y}_u^{T}$ received in $T$ channel uses at the UE and RN are given by deterministic functions of the inputs; that is,
\begin{subequations}
\label{eq:LDM_mod}
\begin{alignat}{2}
\mathbf{y}_r^{T}&=\mathbf{S}^{q-n_{s}}\mathbf{x}_{s}^{T},\label{eq:LDM_mod_1}\\ \mathbf{y}_u^{T}&=\mathbf{S}^{q-n_{d}}\mathbf{x}_{s}^{T}\oplus\mathbf{S}^{q-n_{r}}\mathbf{x}_{r}^{T},\label{eq:LDM_mod_2}
\end{alignat}
\end{subequations} 
where $\mathbf{S}\in\mathbb{F}_2^{q\times q}$ is a down-shift $q\times q$ matrix defined by \begin{equation}\label{eq:shift_mat}
\mathbf{S}=\begin{pmatrix}
\mathbf{0}_{q-1}^{\dagger} & 0 \\ \mathbf{I}_{q-1} & \mathbf{0}_{q-1}
\end{pmatrix}.
\end{equation} The input-output equation \eqref{eq:LDM_mod} approximates the input-output equation of the Gaussian channel given in \eqref{eq:Gaus_mod} and \eqref{eq:Gaus_mod_RN} in the high SNR regime. A graphical representation of the transmitted and received binary vectors $\mathbf{x}_l[i]$ and $\mathbf{y}_j[i]$, $l\in\{r,s\}$, $j\in\{r,u\}$, in the $i$-th channel use is shown in Fig. \ref{fig:ach_1}. 
Each sub-block of length (rate) $R$ in the figure represents a sub-signal being able to hold $R$ bits for transmission. For any wireless link of the network under study, only the most $\mathbf{n}=(n_{d},n_{r},n_{s})^{\dagger}$ significant
bits are received at the destinations while less significant bits are not. We denote the DTB for the LDM by $\Delta_{\text{det}}(\mu,\mathbf{n})$. The remainder of this paper focuses on characterizing the DTB on the basis of the LDM. 

\section{main result}
In this section, we state our main results on the optimal DTB $\Delta^{*}_{\text{det}}(\mu,\mathbf{n})$ for various channel regimes. This results in Theorems \ref{th:1} and \ref{th:2}. To state these theorems we define the following disjoint channel regimes $\{\mathcal{C}_i\}_{i=1}^{4}$: \begin{equation}\label{eq:basic_channel_regimes}
\begin{cases}\mathcal{C}_1=\big\{n_d\geq\max\{n_r,n_s\}\big\}\\\mathcal{C}_2=\big\{n_r \geq n_d\geq n_s\big\}\\\mathcal{C}_3=\big\{\min\{n_r,n_s\}\geq n_d\big\}\\\mathcal{C}_4=\big\{n_s\geq n_d\geq n_r\big\}
\end{cases}.
\end{equation}
First, we provide the following lemma that explains sub-channel regimes derived from $\mathcal{C}_i$ that are utilized in the theorems in greater detail.
\vspace{.5em}
\begin{lemma}\label{lemma}
For the LDM-based cache-aided HetNet in Fig. \ref{fig:HetNet} with $\mu=0$, the optimal DTB is given by \begin{equation}\label{eq:DTB_lemma}
\Delta_{\text{det}}^{*}(\mu=0,\mathbf{n})=\max\bigg\{\frac{2}{\max\{n_{d},n_{s}\}},\frac{1}{n_{s}},\frac{1}{\max\{n_{d},n_r\}}\bigg\}.
\end{equation}
\end{lemma}
\begin{IEEEproof} 
For $\mu=0$, the RN has no relevant information on its own and on the UE's requested file $W_{d_r}$ and $W_{d_u}$. Thus, the DeNB is involved in \emph{broadcasting} files $W_{d_r}$ and $W_{d_u}$ to RN and the UE while the RN acts as a transceiver. The resulting channel constitutes a \emph{broadcast channel with one-sided noisy receiver cooperation}. For a feasible scheme, either user can reliably decode $W_{d_j}$ ($W_{d_r}$ and $W_{d_u}$), $j=\{r,u\}$, if it is aware of $\mathbf{y}_{j}^{T}$ ($\mathbf{S}^{q-\max\{n_d,n_{s}\}}\mathbf{x}_{s}^{T}$). These observations can be used to generate lower bounds on the DTB $\Delta_{\text{det}}^{*}(\mu=0,\mathbf{n})$ that correspond to elements in the outer max-expression of \eqref{eq:DTB_lemma}. Since the requested files are of the same size, an optimal scheme that minimizes the latency would try to split the transmission load equally to the UE and RN. Depending on the channel conditions, the load balancing is done as follows. For instance, when $n_{d}\geq n_{s}$, $\nicefrac{n_{d}}{2}$ bits can be send in one channel use from the DeNB to both RN and UE, if the weaker channel $n_{s}$ is stronger than $\nicefrac{n_{d}}{2}$, i.e., $\nicefrac{n_{d}}{2}\leq n_{s}$. On the other hand if $n_d\geq n_s$ and $\nicefrac{n_d}{2}\geq n_s$, the latency is governed by the weaker channel and $n_{s}$ bits are transmitted in one channel use to each user. In summary, for $n_{d}\geq n_{s}$ we can reliably convey $\min\{\nicefrac{n_{d}}{2},n_{s}\}$ bits per channel use to each user, or in other words, $\Delta^{*}_{\text{det}}(\mu=0,\mathbf{n})=\max\{\nicefrac{2}{n_{d}},\nicefrac{1}{n_{s}}\}$ channel uses are needed to provide each user with one bit. For $n_{d}\leq n_{s}$, a similar observation holds. Details for this case are omitted to conserve space. This concludes the proof.
\end{IEEEproof}
\begin{remark}\label{remark:mu_0} It is easy to verify from Lemma \ref{lemma} that the optimal DTB corresponds to $\nicefrac{2}{n_{d}}$ ($\nicefrac{2}{n_{s}}$) for $\mathcal{I}_0=\big\{2n_{s}\geq n_{d}\geq n_{s}\big\}$ $\big(\mathcal{I}_1=\big\{2\max\{n_{d},n_r\}\geq n_{s}\geq n_{d}\big\}\big)$ and to $\nicefrac{1}{n_{s}}$ ($\nicefrac{1}{\max\{n_{d},n_r\}}$) for $\mathcal{I}_0^{C}=\big\{2n_{s}\leq n_{d}\}$ $\big(\mathcal{I}_1^{C}=\big\{2\max\{n_{d},n_r\}\leq n_{s}\big\}\big)$.
\end{remark}
\vspace{.5em}
Now we are able to state the theorems. We make the distinction between the cases when the direct channel is either weaker or stronger than the DeNB-RN channel ($n_d\geq n_s$ and $n_d\leq n_s$).
\begin{theorem}($n_d\geq n_s$)\label{th:1}
For the LDM-based cache-aided HetNet in Fig. \ref{fig:HetNet} with $\mu\in[0,1]$, the optimal DTB in channel regimes
\begin{equation}\label{eq:channel_regimes}
\begin{cases}\mathcal{R}_1=\mathcal{C}_1\cap \mathcal{I}_0\\\mathcal{R}_1^{\prime}=\mathcal{C}_1\cap \mathcal{I}_0^{C}\\\mathcal{R}_2=\mathcal{C}_2\cap \mathcal{I}_0\\\mathcal{R}_2^{\prime}=\mathcal{C}_2\cap \mathcal{I}_0^{C}\end{cases},
\end{equation} is given by \begin{equation}\label{eq:DTB_theorem}
\Delta_{\text{det}}^{*}(\mu,\mathbf{n})=\begin{cases}\frac{2-\mu}{n_d}\qquad&\text{for }\mathbf{n}\in\mathcal{R}_1\\\max\Big\{\frac{2-2\mu}{n_d},\frac{2-\mu}{n_r}\Big\}\qquad&\text{for }\mathbf{n}\in\mathcal{R}_2\\\max\Big\{\frac{1-\mu}{n_s},\frac{2-\mu}{\max\{n_d,n_r\}}\Big\}\qquad&\text{for }\mathbf{n}\in\mathcal{R}_1^{\prime},\mathcal{R}_2^{\prime}\end{cases}.
\end{equation}
\end{theorem}
\vspace{.5em}
\begin{theorem}($n_d\leq n_s$)\label{th:2}
For the LDM-based cache-aided HetNet in Fig. \ref{fig:HetNet} with $\mu\in[0,1]$, the optimal DTB in the channel regimes 
\begin{equation}
\label{eq:channel_regimes2}
\begin{cases}\mathcal{R}_{31}=\mathcal{C}_3\cap\{n_d\leq n_s\leq 2n_d\}\\\mathcal{R}_{32}=\mathcal{C}_3\cap\{2n_d\leq n_s\leq 2n_r\}\\\mathcal{R}_3^{\prime}=\mathcal{C}_3\cap\mathcal{I}_1^{C}\\\mathcal{R}_4=\mathcal{C}_4\cap \mathcal{I}_1\\\mathcal{R}_4^{\prime}=\mathcal{C}_4\cap\mathcal{I}_1^{C}\end{cases},
\end{equation} corresponds to 
\begin{align}\label{eq:DTB_theorem_2}
\Delta_{\text{det}}^{*}(&\mu,\mathbf{n})\nonumber=\\&\begin{cases}\max\Big\{\frac{2-2\mu}{n_s},\frac{2-\mu}{n_s+(n_r-n_d)^{+}},\Delta_{\text{LB}}(\mathbf{n})\Big\}\qquad&\text{for }\mathbf{n}\in\mathcal{R}_{31}\\\max\Big\{\frac{2-2\mu}{n_s},\Delta_{\text{LB}}(\mathbf{n})\Big\}\qquad&\text{for }\mathbf{n}\in\mathcal{R}_{32}\\\max\Big\{\frac{2-\mu}{n_s+(n_r-n_d)^{+}},\Delta_{\text{LB}}(\mathbf{n})\Big\}\qquad&\text{for }\mathbf{n}\in\mathcal{R}_{4}\\\Delta_{\text{LB}}(\mathbf{n})\qquad&\text{for }\mathbf{n}\in\mathcal{R}_{3}^{\prime},\mathcal{R}_{4}^{\prime}\end{cases},
\end{align}
where $\Delta_{\text{LB}}(\mathbf{n})=\nicefrac{1}{\max\{n_d,n_r\}}$.  
\end{theorem}
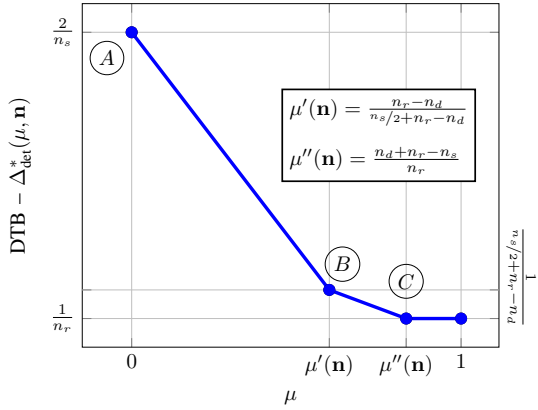
\begin{figure}[!tbp]
  \centering
    \begin{tikzpicture}[scale=0.80]
         \Resultone
    \end{tikzpicture}
    \caption[DTB as a function of $\mu$ for various channel regimes and distinct fronthaul capacities]
    {\small DTB as a function of $\mu$ for $\mathbf{n}\in\mathcal{R}_{31}$} 
    \label{fig:results}
  \end{figure}
\begin{figure*}
        \centering
        \begin{subfigure}[b]{0.425\textwidth}
            \centering
            \begin{tikzpicture}[scale=1]
            \Achievability
            \end{tikzpicture}
           \caption{\small \circled{$B$}$\:=\big(\mu^{\prime}(\mathbf{n}),\frac{1}{\nicefrac{n_s}{2}+n_r-n_d}\big)$ with $\mu^{\prime}(\mathbf{n})=\frac{n_r-n_d}{\nicefrac{n_s}{2}+n_r-n_d}$}    
            \label{fig:ach_1}
        \end{subfigure}
        \hfill
        \begin{subfigure}[b]{0.425\textwidth}  
            \centering 
            \begin{tikzpicture}[scale=1]
            \AchievabilityTwo
            \end{tikzpicture}
           \caption{\small \circled{$C$}$\:=\big(\mu^{\prime\prime}(\mathbf{n}),\frac{1}{n_r}\big)$ with $\mu^{\prime\prime}(\mathbf{n})=\frac{n_d+n_r-n_s}{n_r}$}   
            \label{fig:ach_2}
        \end{subfigure}
        \vskip\baselineskip
                \caption[Achievability Scheme for Two Corner Points] 
        {\small Achievability scheme for corner points (a) \circled{$B$} and (b) \circled{$C$} in channel regime $\mathcal{R}_{31}$ when $n_r\geq n_s$. The length of each sub-block represents the entropy of the corresponding messages. For instance in (a) $H(W_2^{\bar{c}})=\nicefrac{n_s}{2}$.} 
        \label{fig:achievability}
\end{figure*}  
\vspace{.5em}
\noindent \begin{IEEEproof} (Theorems \ref{th:1} and \ref{th:2})
The lower bound (converse) on the DTB is provided in section \ref{sec:lw_bd}. The upper bound (achievability) can be found in section \ref{sec:ach_scheme}. 
\end{IEEEproof}
\vspace{.5em}
\begin{remark}
In all channel regimes of \eqref{eq:channel_regimes} and \eqref{eq:channel_regimes2} (except of $\mathcal{R}_3^{\prime}$ and $\mathcal{R}_4^{\prime}$), caching at the RN decreases the latency with increasing fractional cache size (cf. Fig. \ref{fig:results}) until the best possible latency $\Delta^{*}_{\text{det}}(1,\mathbf{n})=\Delta_{\text{LB}}(\mathbf{n})$ is achieved at some $\mu^{\prime\prime}(\mathbf{n})$. This latency is identical to the optimal DTB when the RN acts as an additional transmitter with access to the entire library of files ($\mu=1$). Interestingly, our results reveal that for $n_d\leq n_s$ only, a fractional cache size $\mu^{\prime\prime}(\mathbf{n})\leq 1$ suffices to attain the exact same performance. For $n_d\geq n_s$, however $\mu^{\prime\prime}(\mathbf{n})= 1$ to achieve a DTB of $\Delta_{\text{LB}}(\mathbf{n})$.         
\end{remark}
\vspace{.5em}
\begin{remark}
In all channel regimes $\mathcal{R}_3^{\prime}$ and $\mathcal{R}_4^{\prime}$ ($\mathcal{R}_3^{\prime}\cup\mathcal{R}_4^{\prime}=\mathcal{I}_1^{C}$), transceiver caching is not beneficial. The optimal latency corresponds to the optimal DTB given in Lemma \ref{lemma} for $\mathbf{n}\in\mathcal{I}_1^{C}$ (see Remark \ref{remark:mu_0}).   
\end{remark}

\section{Lower Bounds on DTB (Converse)}
\label{sec:lw_bd}

In this section, we develop lower bounds on the DTB $\Delta_{\text{det}}(\mu,\mathbf{n})$ to settle the optimality of our proposed achievability scheme for various regimes of channel parameters $n_{d},n_{r}$ and $n_{s}$. For the sake of notational simplicity, we outline the proof as if there were only $N=2$ files. Nonetheless, the lower bound remains valid for $N>2$. We consider one option of a worst-case demand pattern $\mathbf{d}=(d_r, d_u)^{\dagger}=(1,2)^{\dagger}$; that is, RN and UE request \emph{distinct} files $W_{d_r}=W_1$ and $W_{d_u}=W_2$. Throughout this section, we will use $\epsilon_L$ to denote the Fano term with its property $\epsilon_L\rightarrow 0$ as $L\rightarrow\infty$ \cite{Cover_2006}. For a given channel realization $\mathbf{n}=(n_{d},n_{r},n_{s})^{\dagger}$, we establish lower bounds on the delivery time $T$ as the converse on $\Delta_{\text{det}}(\mu,\mathbf{n})$. To this end, we deploy distinct \emph{subsets of information resources} $\mathcal{B}_1,\ldots,\mathcal{B}_5$ that allow for reliable decoding of its corresponding message set as $L\rightarrow\infty$. This is defined as follows:
\begin{align}
&\mathcal{B}_1=\{\mathbf{y}_r^{T},S_1,S_2|W_2\}\rightarrow W_1 \nonumber \\
&\mathcal{B}_2=\{\mathbf{y}_u^{T}\}\rightarrow W_2 \nonumber \\
&\mathcal{B}_3=\{\mathbf{S}^{q-n_d}\mathbf{x}_s^{T},\mathbf{y}_r^{T},S_1,S_2\}\rightarrow W_1,W_2 \nonumber \\
&\mathcal{B}_4=\{\mathbf{y}_u^{T}, S_1\}\rightarrow W_1,W_2\text{ for }n_d\geq n_s
\nonumber \\
&\mathcal{B}_5=\{\mathbf{x}_{s,[n_d+1:n_s]}^{T},\mathbf{y}_u^{T},S_1\}\rightarrow W_1,W_2\text{ for }n_s\geq n_d. \nonumber
\end{align}

For instance, we can write $H(W_1,W_2|\mathcal{B}_5)\leq L\epsilon_L$. As opposed to $\mathcal{B}_2$, subsets $\mathcal{B}_1, \mathcal{B}_3,\mathcal{B}_4$ and $\mathcal{B}_5$ include cached content(s) $S_1$ (and $S_2$) of requested file(s) $W_1$ (and $W_2$). Thus, applying standard information theoretic bounding techniques with subsets $\mathcal{B}$ as side information will generate \emph{cache-dependent} and \emph{independent} bounds through $\mathcal{B}_1,\mathcal{B}_3,\mathcal{B}_4, \mathcal{B}_5$ and $\mathcal{B}_2$, respectively. The reasoning behind the choice of subsets $\mathcal{B}_1$--$\mathcal{B}_4$ are rather intuitive and we omit further details on them for the sake of brevity. We will explain the choice of $\mathcal{B}_5$ in the next paragraph. In accordance with the ordering of the information subsets, when using $\mathcal{B}_1$--$\mathcal{B}_5$ we arrive at the following five inequalities bounding $T$:

{\small
\begin{align}
\label{eq:bound_1} T&\geq \frac{(1-\mu)L}{n_s}-\frac{L\epsilon_L}{n_s},\\ \label{eq:bound_2} T&\geq \frac{L}{\max\{n_d,n_r\}}-\frac{L\epsilon_L}{\max\{n_d,n_r\}}, \\ \label{eq:bound_3} T&\geq \frac{(2-2\mu)L}{\max\{n_d,n_s\}}-\frac{L\epsilon_L}{\max\{n_d,n_s\}}, \\ \label{eq:bound_4} T&\geq \frac{(2-\mu)L}{\max\{n_d,n_r\}}-\frac{L\epsilon_L}{\max\{n_d,n_r\}}, \\ \label{eq:bound_5} T&\geq \frac{(2-\mu)L}{n_s+(n_r-n_d)^{+}}-\frac{L\epsilon_L}{n_s+(n_r-n_d)^{+}}. 
\end{align}
}%
Dividing inequalities \eqref{eq:bound_1}--\eqref{eq:bound_5} by $L$ and taking the limit for $L\rightarrow\infty$ will yield the desired lower bounds. 

In the remainder of this section, we will explain the intuition behind the choice of $\mathcal{B}_5=\{\mathbf{x}_{s,[n_d+1:n_s]}^{T},\mathbf{y}_u^{T},S_1\}$ applicable for $n_s\geq n_d$ only. We note that knowing $\mathbf{y}_r^{T}$ on top of $\mathcal{B}_5$ suffices for reliable decodability. In fact, we will show that $\mathbf{y}_r^{T}$ can be directly recovered from $\mathcal{B}_5$. To understand this fact, we make two key observations:   
\begin{enumerate}[label=(\Alph*)]
\item First, we note that through the channel output $\mathbf{y}_u^{T}\subseteq \mathcal{B}_5$ spanning $T$ channel uses, any hypothetical decoder obtains file $W_2$ and ultimately $S_2$ since $S_2=\phi_2(W_2)$ ($\mathbf{y}_u^{T}\rightarrow W_2\rightarrow S_2$).
\item Second, any decoder can retrieve $\mathbf{y}_r^{T}$ from $\{\mathbf{x}_{s,[n_d+1:n_s]}^{T},\mathbf{y}_u^{T},S_1,S_2\}$ in a \emph{recursive} manner.  
\end{enumerate} \vspace{0.5em}
Note that (B) presumes observation (A) due to the awareness of $S_2$ from $\mathbf{y}_u^{T}$. Key observation (B) can be explained thoroughly through $T$ recursive steps $t=1,\ldots, T$. In this context, the terms steps and channel use are equipollent. We now describe the main operations of step $t=1$ in detail. First, all arguments of RN's encoding function for channel use $t=1$ are readily available allowing for the recovery of $\mathbf{x}_r[1]$ according to
\begin{equation}
\label{eq:relay_signal_time_1}
\mathbf{x}_r[1]=\psi_{r}^{[1]}(S_1,S_2,\mathbf{y}_r[0]=\emptyset,\mathbf{d},\mathbf{h}).
\end{equation} Next, one can obtain the top most $n_d$ samples of $\mathbf{x}_s[1]$ specified by $\mathbf{x}_{s,[1:n_d]}[1]$ from the known vectors $\mathbf{x}_r[1]$ and $\mathbf{y}_u[1]$ in Equation \eqref{eq:LDM_mod_2}. Concatenating $\mathbf{x}_{s,[1:n_d]}[1]$ and $\mathbf{x}_{s,[n_d+1:n_s]}[1]$ to a single vector generates $\mathbf{x}_{s}[1]$; thus, also $\mathbf{y}_{r}[1]$ (cf. \eqref{eq:LDM_mod_1}). With $\mathbf{y}_{r}[1]$ in place, the RN encoding function $\psi_{r}^{[2]}$ at channel use $t=2$ can be invoked, and, therefore step $t=2$ initiates. The decoder proceeds this way in the remaining $T-1$ steps until all instances (up to the $T$--th instance) of $\mathbf{y}_{r}^{T}$ have been generated. Finally, now having $\mathbf{y}_{r}^{T}\cup\mathcal{B}_5$, decoding of $W_1$ and $W_2$ becomes possible. On the example of $\mathcal{B}_5$, the main bounding techniques applied to establish the lower bounds look as follows:
\begin{align}\label{eq:conv_1}
2L=\:\:&H\big(W_1,W_2\big)\nonumber \\ =\:\:& I\big(W_1,W_2;\mathbf{x}_{s,[n_d+1:n_s]}^{T},\mathbf{y}_u^{T},S_1\big)\nonumber\\\:\:&+H\big(W_1,W_2|\mathbf{x}_{s,[n_d+1:n_s]}^{T},\mathbf{y}_u^{T},S_1\big)\nonumber\\\stackrel{(a)} \leq\:\:& H\big(\mathbf{x}_{s,[n_d+1:n_s]}^{T}\big)+H\big(\mathbf{y}_u^{T}\big)+H\big(S_1\big)+L\epsilon_L
\nonumber\\ \stackrel{(b)} \leq\:\:& T\big(n_s+(n_r-n_d)^{+}\big)+\mu L+L\epsilon_L,
\end{align} 
where (a) follows from Fano's inequality and the fact that conditioning does not increase entropy and (b) is because the $\text{Bern}(\nicefrac{1}{2})$ distribution maximizes the binary entropy of each element of $\mathbf{x}_{s,[n_d+1:n_s]}^{T}$ and $\mathbf{y}_u^{T}$ as well as $H\big(S_1\big)\leq\mu L$.

\section{Achievability of Minimum DTB}
\label{sec:ach_scheme}
In this section, we provide details on our scheme that achieves the minimum DTB. Due to page limitations, the scheme is described for the single channel regime $\mathcal{R}_{31}$. Details on the achievability for the remaining channel regimes follow similar lines of argument. We first observe that the minimum DTB is a convex function of the fractional cache size $\mu$. 
\vspace{.5em}
\begin{lemma}\label{lemma:2}
The minimum DTB $\Delta^{*}_{\text{det}}(\mu,\mathbf{n})$ is a convex function of $\mu\in[0,1]$ for any given channel $\mathbf{n}$.
\end{lemma}
\noindent\begin{IEEEproof} The proof is omitted due to lack of space.       
\end{IEEEproof}
\vspace{.5em} 
Due to the convexity of the DTB, we infer that it suffices to establish the achievability of the corner points \circled{$A$},\circled{$B$} and \circled{$C$} (cf. Fig. \ref{fig:results}). In fact, the optimal DTB at fractional cache size $\mu$ which lies between two neighboring corner points (say \circled{$A$} and \circled{$B$} for instance) cache sizes' is achieved through file splitting and time sharing between the policies at those two corner points. This strategy is only operational at a channel regime under which the transmission policy at \circled{$A$}, $\mathcal{R}_{A}$, and the transmission policy at \circled{$B$}, $\mathcal{R}_{B}$, are \emph{both} feasible. This is given by the \emph{non-empty} set $\mathcal{R}_{A}\cap\mathcal{R}_{B}$. The next two sub-sections establish the achievability for \circled{$A$}, \circled{$B$} and \circled{$C$}, respectively. In either sub-section, it is assumed that the RN and the UE request different files denoted by $W_1$ and $W_2$. 
\subsection{Achievability at $\mu=0$}
For $\mu=0$, we can directly apply the results of Lemma \ref{lemma}. For $\mathcal{R}_{A}=\mathcal{I}_1\supseteq \{n_d\leq n_s\leq 2n_d\}$, this establishes the achievability of a $\nicefrac{2}{n_s}$-DTB.  
\subsection{Achievability at $\mu^{\prime}(\mathbf{n})$ and $\mu^{\prime\prime}(\mathbf{n})$}
The optimal schemes at both $\mu^{\prime}(\mathbf{n})$ and $\mu^{\prime\prime}(\mathbf{n})$ operate over one single channel use under the same channel regime $\mathcal{R}_{B}=\mathcal{R}_{C}=\mathcal{R}_{31}$. Both transmission schemes are very much alike as they are based upon the same premise. That is, the DeNB transmits parts of the requested files $W_1$ and $W_2$ that are \emph{not} cached ($W_1^{\bar{c}}$ and $W_2^{\bar{c}}$) at the RN. The RN, on the other hand, applies receiver caching for its own sake and transmitter caching for the sake of the UE to make the remaining, \emph{cached} parts of $W_1$ and $W_2$ ($W_1^{c}$ and $W_2^{c}$) available. As $\mu$ increases, more information on the requested files are available locally at the RN; thus, reducing, the required rates on $W_1^{\bar{c}}$ and $W_2^{\bar{c}}$. The proper rate allocation of $W_i^{\bar{c}}$ and $W_i^{c}$, $i=1,2$, at $\mu^{\prime}(\mathbf{n})$ and $\mu^{\prime\prime}(\mathbf{n})$ ($\mu^{\prime}(\mathbf{n})\leq\mu^{\prime\prime}(\mathbf{n})$) is specified in Figs. \ref{fig:ach_1} and \ref{fig:ach_2}, respectively. The required fractional cache size at these corner points corresponds to 
\begin{equation}
\label{eq:req_ach_cache_size}
\mu=\frac{H(W_i^{c})}{H(W_i)},\qquad i=1,2.
\end{equation} 
As the transmission schemes need \emph{one channel use} to convey transceiver RN and the UE with files $W_i$ of equal rate, the achievable DTB becomes the inverse of the file rate, i.e., $\nicefrac{1}{L}$, where $L=H(W_i^{c})+H(W_i^{\bar{c}})$. This establishes the achievability for corner points \circled{$B$} and \circled{$C$}.

\section{Conclusion}
\label{sec:conlusion}

In this paper, we have studied the fundamental latency of the transceiver cache-aided wireless HetNet in the downlink given in Fig. \ref{fig:HetNet}. We utilize the DTB as the performance metric that captures the worst-case delivery latency of requested files. The LDM is used as an approximation method for Gaussian channels, to completely characterize the optimal tradeoff between storage and latency for the given HetNet in various channel regimes. 

\bibliographystyle{IEEEtran}
\bibliography{content/bibliography}
\balance
 
\end{document}